\documentclass[%
 reprint,
]{revtex4-2}

\usepackage{physics}
\usepackage{graphicx}
\usepackage{dcolumn}
\usepackage{bm}
\usepackage{appendix}
\usepackage{amsfonts}
\usepackage[normalem]{ulem}
\usepackage{xcolor}

\begin{document}

\title{Bounds on an effective thermalization beyond the Zeno limit}
\author{Guilherme Zambon}
\email{guilhermezambon@usp.br}
\author{Diogo O. Soares-Pinto}
\email{dosp@ifsc.usp.br}
\affiliation{
 Instituto de F{\'i}sica de S{\~a}o Carlos, Universidade de S{\~a}o Paulo, CP 369, 13560-970, S{\~a}o Carlos, SP, Brasil.
 }

\begin{abstract}
Developing protocols for preserving information in quantum systems is a central quest for implementing realistic quantum computation. In this regard, the quantum Zeno effect has emerged as a widely utilized technique to safeguard classical information stored in quantum systems. However, existing results pertaining to this method often assume operations performed infinitely fast on the system of interest, which only serves as an approximation to real-world scenarios where the temporal resolution of any experimental apparatus is inherently finite. In this study, we go beyond this conventional assumption and derive the effective Zeno dynamics for any time interval between operations. Our analysis considers a qubit undergoing thermalization, as described by a generalized amplitude damping channel, while the operations performed consist of projections onto an orthonormal basis that may or may not coincide with the pointer basis to which the system is thermalizing. By obtaining the probability of successfully storing a bit of information after a given time, we investigate the performance of the protocol in two important scenarios: the limit of many interventions, with a first-order correction to the Zeno limit, and the limit of very few interventions. In doing so, we provide valuable insights into the protocol's performance by establishing bounds on its efficacy. These findings enhance our understanding of the practical applicability of the quantum Zeno effect in preserving classical information stored in quantum systems, allowing for better design and optimization of quantum information processing protocols.
\end{abstract}

\maketitle

\section{\label{sec:introduction}Introduction}

Building a universal quantum computer has been so far one of the most challenging tasks of the twenty-first century Physics. Since Feynman's famous insight on employing controlled quantum systems to simulate other quantum systems found in nature \cite{feynman2018simulating}, physicists have been developing theoretical and experimental tools to make these simulations possible. Ideally, everything that is needed to perform useful quantum computation is a set of physical qubits and the ability to experimentally implement any sequence of logical gates on this set. In reality, however, this might not be enough.

As no quantum system is ever truly isolated from its surroundings, during the computation the evolution of the experimentally controlled physical systems might be affected by the interactions between them and their environment \cite{rivas2012open}. Such effects, generally known as noise, may continuously modify the quantum state stored in the qubits, which is problematic, as it is essential to preserve this state during and in between computational steps to reliably implement any algorithm on a real quantum computer.

To address these challenges, techniques such as quantum error correction and dynamical decoupling can be employed. While these methods offer valuable approaches for protecting quantum states, they have inherent drawbacks. Quantum error correction requires storing an informational qubit in multiple physical qubits, necessitating substantial overhead and resources \cite{terhal2015quantum,shor1995scheme,laflamme1996perfect,calderbank1996good,steane1996error,freedman2003topological,brun2006correcting,pastawski2015holographic,fiusa2022fidelity,boyers2019floquet}. On the other hand, dynamical decoupling relies on precise knowledge of the qubit-environment dynamics, which may be challenging to ascertain in realistic scenarios \cite{viola1998dynamical,viola1999dynamical,viola2003robust,khodjasteh2005fault,uhrig2007keeping,west2010high,yang2011preserving,khodjasteh2013designing,du2009preserving,szwer2010keeping,biercuk2009experimental,souza2012robust,zhang2015experimental,zurek1981pointer}.

Alternatively, if our goal is solely to preserve classical information stored in a quantum system, the quantum Zeno effect presents a compelling approach, as it circumvents the drawbacks of the aforementioned techniques \cite{itano1990quantum,presilla1996measurement,kondo2016using,mobus2019quantum,burgarth2019generalized,burgarth2020quantum,becker2021quantum}. It consists in performing repeated projective measurements on the system of interest, which in the limit of infinitely frequent operations (Zeno limit) is known to freeze the dynamics at some subspaces and to generate an effective Zeno dynamics at others \cite{alvarez2006environmentally,alvarez2007decoherence,fernandez2015decoherent}.

Given a free dynamics and a set of operations to be periodically performed, one could obtain the effective dynamics in the Zeno limit using a suitable result, like those found in Refs. \cite{mobus2019quantum,burgarth2019generalized,burgarth2020quantum,becker2021quantum}. However, in practical experimental scenarios there is always a limitation on the frequency at which external operations can be realized. This limitation implies that while the Zeno limit serves as a useful approximation in certain cases, it is ultimately unattainable.

Here, we address this limitation by explicitly determining the effective dynamics for any time interval between operations for the relevant case of a qubit undergoing thermalization, as described by a generalized amplitude damping channel. The operations performed are non-selective projective measurements in as basis that may or may not coincide with the pointer basis to which the qubit is thermalizing. We particularly explore the Zeno limit, a first order correction to it and also the limit of very few operations performed. Finally, these results are used to establish bounds on the protocol's efficiency, as quantified by the probability of successfully storing one bit of classical information on a thermalizing qubit for some given period of time.

\section{\label{sec:setup}Setup}

Consider a qubit used to store a classical bit of information over some period of time. Given its interaction with the environment, it must be treated as an open quantum system, whose state evolves non-unitarily \cite{rivas2012open}. A common approach to this situation is to model the effects of the environment with a quantum channel $\mathcal{E}_t$, such that the state of the qubit at some time $t$ will be given by
\begin{equation}
\label{eq:operator-sum}
    \begin{aligned}
    \rho_t=&~\mathcal{E}_t(\rho_{\mathrm{in}})\\
    =&\sum_{i=0}^3K_i(t)\rho_{\mathrm{in}}K_i^\dagger(t),
    \end{aligned}
\end{equation}
where $K_i(t)$ are the Kraus operators associated to the channel $\mathcal{E}_t$ and $\rho_{\mathrm{in}}$ is the initial state of the qubit \cite{wilde2013quantum}.

For many relevant situations, the environment can be simply modeled as a thermal bath at some fixed temperature $T$, so the qubit is expected to thermalize at this temperature and asymptotically approach a thermal state. This type of evolution is achieved by using the amplitude damping channel $\mathcal{E}_{AD}$ \cite{nielsen2002quantum}, whose Kraus operators are given by
\begin{subequations}
\label{eq:kraus}
    \begin{equation}
        K_0=\sqrt{p}\qty(\ket{e_0}\bra{e_0}+\sqrt{1-\gamma}\ket{e_1}\bra{e_1}),
    \end{equation}
    \begin{equation}
        K_1=\sqrt{p\gamma}\ket{e_0}\bra{e_1},
    \end{equation}
    \begin{equation}
        K_2=\sqrt{1-p}\qty(\sqrt{1-\gamma}\ket{e_0}\bra{e_0}+\ket{e_1}\bra{e_1}),
    \end{equation}
    \begin{equation}
        K_3=\sqrt{(1-p)\gamma}\ket{e_1}\bra{e_0}.
    \end{equation}
\end{subequations}

In the above equations, $\gamma:=\gamma_t$ is a temporal parameter monotonically increasing in $t$, satisfying $\gamma_0=0$ and $\gamma_t\to 1$ as $t\to\infty$. It can be verified that $\gamma_t=0$ implies $\mathcal{E}_{AD}(\rho)=\rho$, as expected, while $\gamma_t=1$ implies $\mathcal{E}_{AD}(\rho)=p\ket{e_0}\bra{e_0}+(1-p)\ket{e_1}\bra{e_1}$, which is a thermal state in the basis $\{\ket{e_0},\ket{e_1}\}$. It follows that $p=p(T)$ is the population of the ground state in thermal equilibrium. Setting its energy to be zero and the energy of the excited state to be $E$, yields
\begin{equation}
    p(T)=\frac{1}{1+e^{-E/k_BT}},
\end{equation}
implying $p\to1$ as $T\to0$ and $p\to1/2$ as $T\to\infty$.

Notice that in thermal equilibrium the system will be diagonal in the \textit{pointer basis} $\{\ket{e_0},\ket{e_1}\}$, determined by the system-environment dynamics \cite{zurek1981pointer}. Nevertheless, we assume the experimenter could choose any other basis to be the \textit{computational basis} $\{\ket{0},\ket{1}\}$ in which the classical bit is initially stored, either as $\rho_{\mathrm{in}}=\ket{0}\bra{0}$ or as $\rho_{\mathrm{in}}=\ket{1}\bra{1}$. Each $\ket{j}$ can still be written in the pointer basis using a convenient parametrization,
\begin{subequations}
    \begin{equation}
        \ket{0}=\cos\theta\ket{e_0}+e^{i\phi}\sin\theta\ket{e_1},
    \end{equation}
    \begin{equation}
        \ket{1}=\sin\theta\ket{e_0}-e^{i\phi}\cos\theta\ket{e_1},
    \end{equation}
\end{subequations}
for $\theta\in[0,\pi/2]$ and $\phi\in[0,2\pi)$. In  Appendix \ref{sec:app-ortho} we show that this is indeed the most general form of the computational basis, given the orthogonality constraint $\bra{0}\ket{1}=0$.

To analyze the dynamics of the system we must propose an explicit expression for $\gamma_t$, which depends on the specific physical situation that is being represented by the amplitude damping channel. Here, we assume that the bath is much larger than the qubit and also that their coupling is sufficiently weak, implying that there are no memory effects due to information backflow, i.e., the dynamics of the qubit is Markovian \cite{rivas2014quantum,breuer2016colloquium,de2017dynamics,li2018concepts,chruscinski2022dynamical}. This is done by imposing that the family of quantum channels is CP-divisible and homogeneous in time 
\begin{equation}
    \label{eq:homogeneous}
    \mathcal{E}_{t+t^\prime}=\mathcal{E}_{t^\prime}\mathcal{E}_t.
\end{equation}
In Appendix \ref{sec:app-gamma} it is shown that this condition implies an exponential decay for $\gamma_t$, that is
\begin{equation}
    \gamma_t=1-e^{-t/\tau},
\end{equation}
where $\tau$ is some characteristic relaxation time determined by the system-environment interaction dynamics \cite{nielsen2002quantum}.

Consider the qubit starts in a pure state in the computational basis $\rho_{\mathrm{in}}=\ket{j}\bra{j},~j=0,1$. It is then subject to non-selective projective measurements in this basis after every interval $\Delta$ of time, which can be seen as the temporal resolution of the experimental apparatus. In the Zeno limit we have $\Delta\to0$, while in the limit of few measurements after some time $t$ we have $\Delta\sim t$. We define the ground state population after some time $t$ as
\begin{equation}
    a(t):=\bra{0}\rho_t\ket{0},
\end{equation}
with $a_0:=a(0)=1-j$. Now we determine how this population evolves in time.

\section{\label{sec:results}Results and Discussion}

\subsection{\label{subsec:dynamics}Effective dynamics}

Between measurements, the state of the qubit will simply evolve according to the amplitude damping channel, exponentially approaching the thermal state on the pointer basis. This implies the population $a(t)$ will change and the coherences $\bra{0}\rho_t\ket{1}$ and $\bra{1}\rho_t\ket{0}$ will become non-zero. However, when the next measurement is performed these coherences will vanish, while the populations will remain unchanged. Therefore, the time dependence of $a(t)$ will be determined by the probabilities of the system transitioning from one state to another within an interval $\Delta$ of time. We use the conditional probability rule to define the transitioning probabilities
\begin{equation}
\label{eq:prob}
    \begin{aligned}
        a(t+\Delta)=&~P(0,t+\Delta)\\
        =&~P(0,t)P(0,t+\Delta|0,t)\\
        +&~P(1,t)P(0,t+\Delta|1,t)\\
        =&~a(t)P_{0|0}+(1-a(t))P_{0|1},
    \end{aligned}
\end{equation}
where we have defined $P_{0|1}$ as the probability of the qubit transitioning from $1$ to $0$ within an interval $\Delta$ of time, and the normalization implies $P_{0|0}=1-P_{1|0}$.

Notice that due to the Markov hypothesis, the transition probabilities $P_{1|0}$ and $P_{0|1}$ are only functions of $\Delta$ and not of $t$. Eq. \eqref{eq:prob} then yields
\begin{equation}
    \begin{aligned}
    \label{eq:recurrence}
        a(t+\Delta)=&~P_{0|1}+a(t)\qty[1-\qty(P_{0|1}+P_{1|0})].
    \end{aligned}
\end{equation}
In Appendix \ref{sec:app-ground} we recurrently expand the above equation from $a_0$, resulting in the following ground state population for any $t=n\Delta$
\begin{equation}
    a(t)=a_{\infty}-\qty(a_{\infty}-a_0)\qty[1-\qty(P_{0|1}+P_{1|0})]^{t/\Delta},
\end{equation}
where
\begin{equation}
    \begin{aligned}
        a_{\infty}:=&~\lim_{t\to\infty}a(t)\\
        =&~\frac{P_{0|1}}{P_{0|1}+P_{1|0}}
    \end{aligned}
\end{equation}
is the (computational basis) ground state population at thermal equilibrium.

In Appendix \ref{sec:app-trans}, we explicitly determine the transition probabilities as functions of $\Delta$ and $\theta$, which allows us to write
\begin{equation}
\label{eq:general}
    \begin{aligned}
        a(t,\Delta,\theta)=&~a_{\infty}-\qty(a_{\infty}-a_0)\Big[\sin^2(2\theta)e^{-\Delta/2\tau}\\
        +&~\cos^2(2\theta)e^{-\Delta/\tau}\Big]^{t/\Delta}
    \end{aligned}
\end{equation}
and
\begin{equation}
    a_{\infty}(\Delta,\theta)=\frac{1}{2}\qty[1+\frac{(2p-1)\cos(2\theta)}{\cos^2(2\theta)+\sin^2(2\theta)\frac{1-e^{-\Delta/2\tau}}{1-e^{-\Delta/\tau}}}].
\end{equation}

It should be pointed out that this expression for the ground state population is only exact for instants right after the measurement is performed, that is, for $t=n\Delta$. With this result in hand, we are able to explore some relevant limits regarding $\Delta$.

\subsubsection{\label{subsubsec:zeno_limit}Zeno limit}

We first define dimensionless quantity $\delta:=\Delta/\tau$. The Zeno limit is then given by $\delta\to0$, which means the interval between measurements is much smaller than the characteristic decay time of the thermalization,
\begin{equation}
\label{eq:zeno}
    \begin{aligned}
        a^{\mathrm{Zeno}}(t,\theta):=&\lim_{\delta\to0} a(t,\delta,\theta)\\
        =&~a_{\infty}^0-e^{-t/\tau^{\mathrm{eff}}}(a_{\infty}^0-a_0),
    \end{aligned}
\end{equation}
where
\begin{equation}
    \begin{aligned}
        a_{\infty}^0(\theta):=&\lim_{\delta\to0} a_{\infty}(\delta,\theta)\\
        =&~\frac{1}{2}+\frac{(2p-1)\cos(2\theta)}{1+\cos^2(2\theta)}
    \end{aligned}
\end{equation}
is the ground state population at thermal equilibrium in the Zeno limit (or the zeroth order expansion of $a_{\infty}$ in $\delta$), and
\begin{equation}
    \tau^{\mathrm{eff}}(\theta):=\frac{4\tau}{3+\cos(4\theta)},
\end{equation}
is the effective characteristic time of the thermalization.

\begin{figure}
    \centering
    \includegraphics[width=\columnwidth]{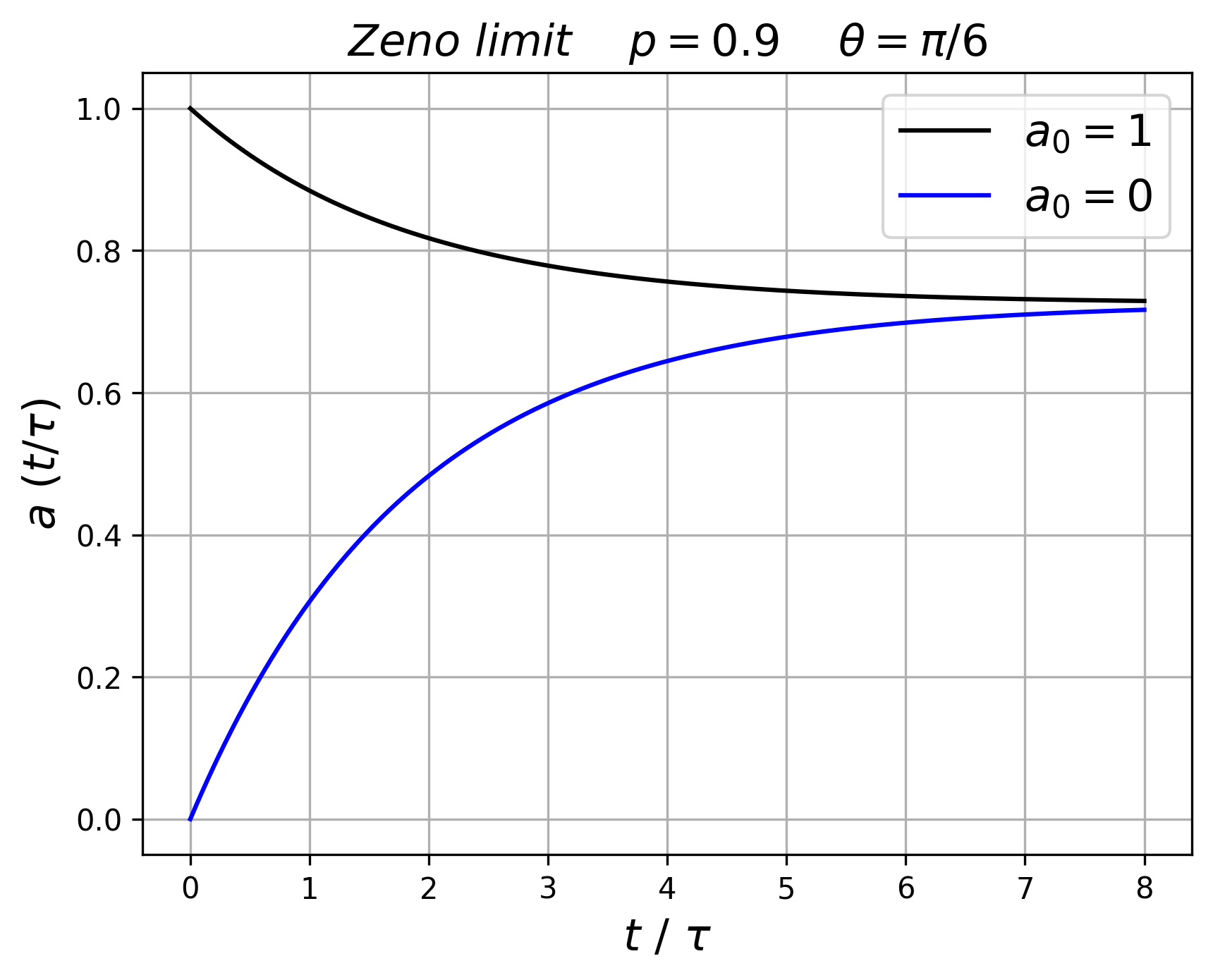}
    \caption{Computational basis ground state population as a function of time. The graph shows the specific case in which the angle $\theta$ between $\ket{e_0}$ and $\ket{0}$ is $\pi/6$ and the $\ket{e_0}$ state population at thermal equilibrium is $p=0.9$, implying the temperature of the environment is relatively low. The $\ket{0}$ state population population exponentially approaches its equilibrium value, which is different from $p$. This happens for both $\rho_{\mathrm{in}}=\ket{0}\bra{0}$ ($a_0=1$) and $\rho_{\mathrm{in}}=\ket{1}\bra{1}$ ($a_0=0$), that is, independently of the classical bit being initially stored.}
    \label{fig:zeno}
\end{figure}

As in this limit the system is never able to create coherences in the computational basis, the dynamics is fully described $a^{\mathrm{Zeno}}(t,\theta)$. Also, this expression is valid for any $t$. As we can see in Fig. \ref{fig:zeno}, the dynamics is a single exponential starting in $a_0$ and asymptotically approaching $a_{\infty}$, with characteristic time $\tau^{\mathrm{eff}}$. Such behavior can be described as an effective thermalization, as the state is constrained by the measurements to remain diagonal in the computational basis and exponentially approaches an effective thermal state. As in general $a_{\infty}\ne p$, we might have an effective temperature which differs from the original one. Importantly, given that $\tau^{\mathrm{eff}}\ge\tau$, the effective thermalization is always slower than the original one, implying that the information about the initial state is kept for longer times in the Zeno effective dynamics. 

\subsubsection{\label{subsubsec:first}First order correction}

In situations where the approximation $\delta\sim 0$ is not a good one, a first order correction to the Zeno limit might be useful. We define $c^1$ as the first order correction in $\delta$ to $a(t,\delta,\theta)$,
\begin{equation}
    a(t,\delta,\theta) = a^{\mathrm{Zeno}}(t,\theta) + \delta~c^1(t,\theta) + \mathcal{O}(\delta^2).
\end{equation}
Performing the expansion of \eqref{eq:general} up to first order in $\delta$ yields the correction
\begin{equation}
    c^1(t,\theta) = c_{\infty}^1-e^{-t/\tau^{\mathrm{eff}}}(c_{\infty}^1-c_0^1),
\end{equation}
with
\begin{equation}
    c_{\infty}^1(\theta) = -\frac{(2p-1)\cos(2\theta)\sin^2(2\theta)}{\qty[3+\cos(4\theta)]^2}
\end{equation}
and
\begin{equation}
    c_0^1(t,\theta) = a_0\frac{\sin^2(4\theta)}{32}t.
\end{equation}
Alternatively, we could explicitly write $a(t,\delta,\theta)$ up to first order in $\delta$ as
\begin{equation}
    \begin{aligned}
        a^1(t,\delta,\theta)=a_{\infty}^1-e^{-t/\tau^{\mathrm{eff}}}(a_{\infty}^1-a_0^1),
    \end{aligned}
\end{equation}
where
\begin{equation}
    \begin{aligned}
        a_{\infty}^1(\theta,\delta)=&~\frac{1}{2}+\frac{2(2p-1)\cos(2\theta)}{3+\cos(4\theta)}\\
        -&~\delta\frac{(2p-1)\cos(2\theta)\sin^2(2\theta)}{\qty[3+\cos(4\theta)]^2}
    \end{aligned}
\end{equation}
is the first order expansion of $a_{\infty}$ in $\delta$, and
\begin{equation}
\label{eq:linear}
    a_0^1(t,\delta,\theta):=a_0\qty(1+\delta \frac{\sin^2(4\theta)}{32}t).
\end{equation}

\begin{figure}
    \centering
    \includegraphics[width=\columnwidth]{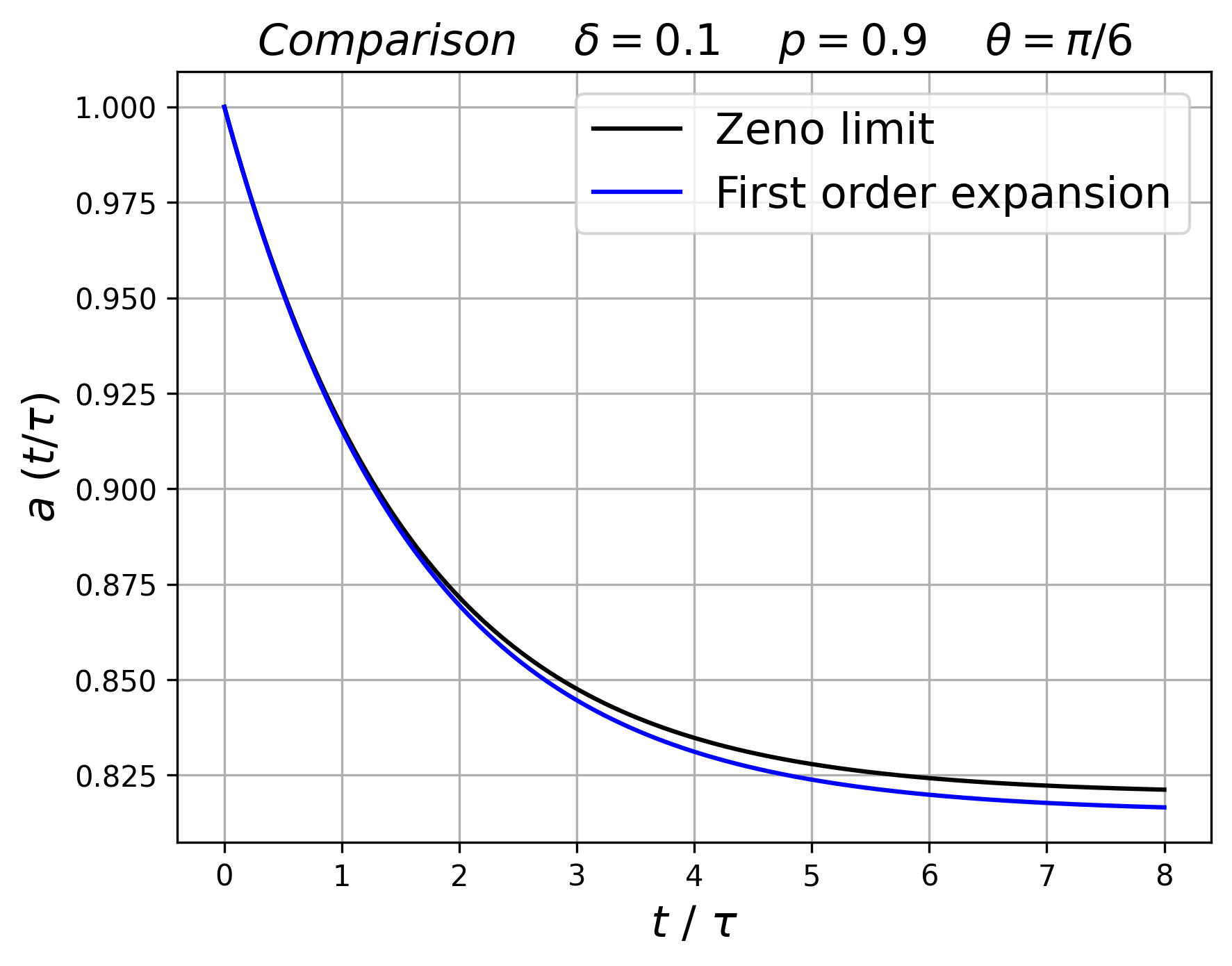}
    \caption{Comparison between the $\ket{0}$ state population in the Zeno limit and its expansion up to first order in $\delta$ for the case $\delta=0.1$. This shows how the system evolves differently when the approximation $\delta\sim0$ ceases to be a good one. In this case, the characteristic time $\tau$ of the thermalization is only $10$ times greater than the time $\Delta$ between measurements.}
    \label{fig:first}
\end{figure}

Interestingly, the first order correction to $a(t,\delta,\theta)$ does not change the characteristic thermalization time $\tau^{\mathrm{eff}}$ and provides a first order correction to the equilibrium value $a_{\infty}$. However, the dynamics will not be described by a simple exponential anymore, as there will be also a correction to the initial state which is not constant, but linear in time, as shown in Eq. \eqref{eq:linear}. The comparison between the Zeno limit and the first order dynamics is shown in Fig. \ref{fig:first}.

\subsubsection{\label{subsubsec:free}Free limit}

\begin{figure}
    \centering
    \includegraphics[width=\columnwidth]{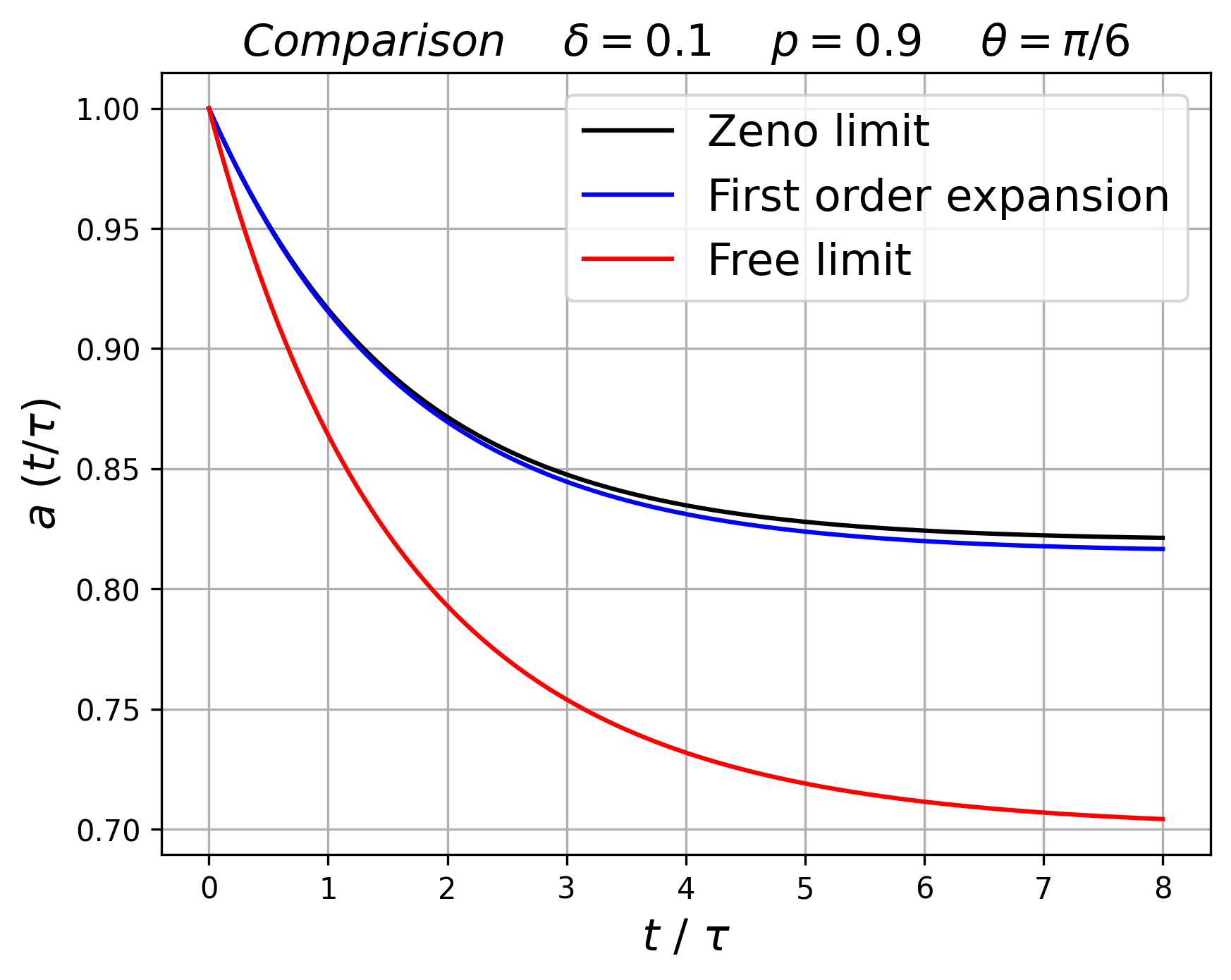}
    \caption{Comparison between the Zeno limit and the Free limit of the evolution. In this particular case they differ substantially. However, comparing Eqs. \eqref{eq:zeno} and \eqref{eq:free} we can see that they coincide when the angle $\theta$ is $0$, $\pi/4$ or $\pi/2$.}
    \label{fig:free}
\end{figure}

On the other side of the $\delta$ limits we have the nearly-free dynamics, where very few measurements are performed within a period of time. Since our general expression for the ground state population is only exact for $t=n\Delta$, and we are treating the case $\Delta\sim t$, it is convenient to consider $\Delta=t$, which is the case in which we leave the qubit evolve freely during the whole time $t$, and only measure it in the end of the storage time. Eq. \eqref{eq:general} then yields
\begin{equation}
\label{eq:free}
    \begin{aligned}
        a^{\mathrm{Free}}(t,\theta)=&~a_{\infty}^F-\qty(a_{\infty}^F-a_0)\Big[\sin^2(2\theta)e^{-t/2\tau}\\        
        +&~\cos^2(2\theta)e^{-t/\tau}\Big],
    \end{aligned}
\end{equation}
where
\begin{equation}
    a_{\infty}^F(\theta) = \frac{1}{2}\qty[1+\qty(2p-1)\cos(2\theta)].
\end{equation}
Notice that in this case the dynamics is described by a combination of two exponentials, one decaying with $\tau$ and the other with $2\tau$. The comparison between the free limit and the Zeno limit is shown in Fig. \ref{fig:free}.

\subsection{\label{subsec:psuc}Success probability}

Now that we have determined how the ground state population of the pointer basis evolve in several situations, we proceed to the calculation of the probability of successfully storing the bit of classical information within a period of time. First, notice that the probability of a successful storage given that we started with $0$ is simply the ground state population with $a_0=1$. On the other hand, the probability of a successful storage given that we started with $1$ is the excited state population, which is given by $(1-a)$ with $a_0=0$. Then, assuming the bits $0$ and $1$ are equally likely to be stored at this point of the computation, we obtain
\begin{equation}
\label{eq:psuc}
    P_{\mathrm{suc}}=~\frac{1}{2}\qty[a_{a_0=1}+(1-a_{a_0=0})].
\end{equation}

Combining this equation with the expressions for the ground state population we obtain for a general $\Delta$
\begin{equation}
    \begin{aligned}
        P_{\mathrm{suc}}=&~\frac{1}{2}\Bigg\{1+\Big[\sin^2(2\theta)e^{-\Delta/2\tau}+\cos^2(2\theta)e^{-\Delta/\tau}\Big]^{t/\Delta}\Bigg\},
    \end{aligned}
\end{equation}
while for the Zeno limit
\begin{equation}
    \begin{aligned}
        P_{\mathrm{suc}}^{\mathrm{Zeno}}=&~\frac{1}{2}\qty[1+e^{-t/\tau^{\mathrm{eff}}}],
    \end{aligned}
\end{equation}
the first order expansion
\begin{equation}
    \begin{aligned}
        P_{\mathrm{suc}}^1=&~\frac{1}{2}\qty[1+e^{-t/\tau^{\mathrm{eff}}}\qty(1+\delta \frac{\sin^2(4\theta)}{32}t)],
    \end{aligned}
\end{equation}
and the free limit
\begin{equation}
    \begin{aligned}
        P_{\mathrm{suc}}^{\mathrm{Free}}=&~\frac{1}{2}\qty[1+\sin^2(2\theta)e^{-t/2\tau}+\cos^2(2\theta)e^{-t/\tau}].
    \end{aligned}
\end{equation}

\begin{figure}
    \centering
    \includegraphics[width=0.945\columnwidth]{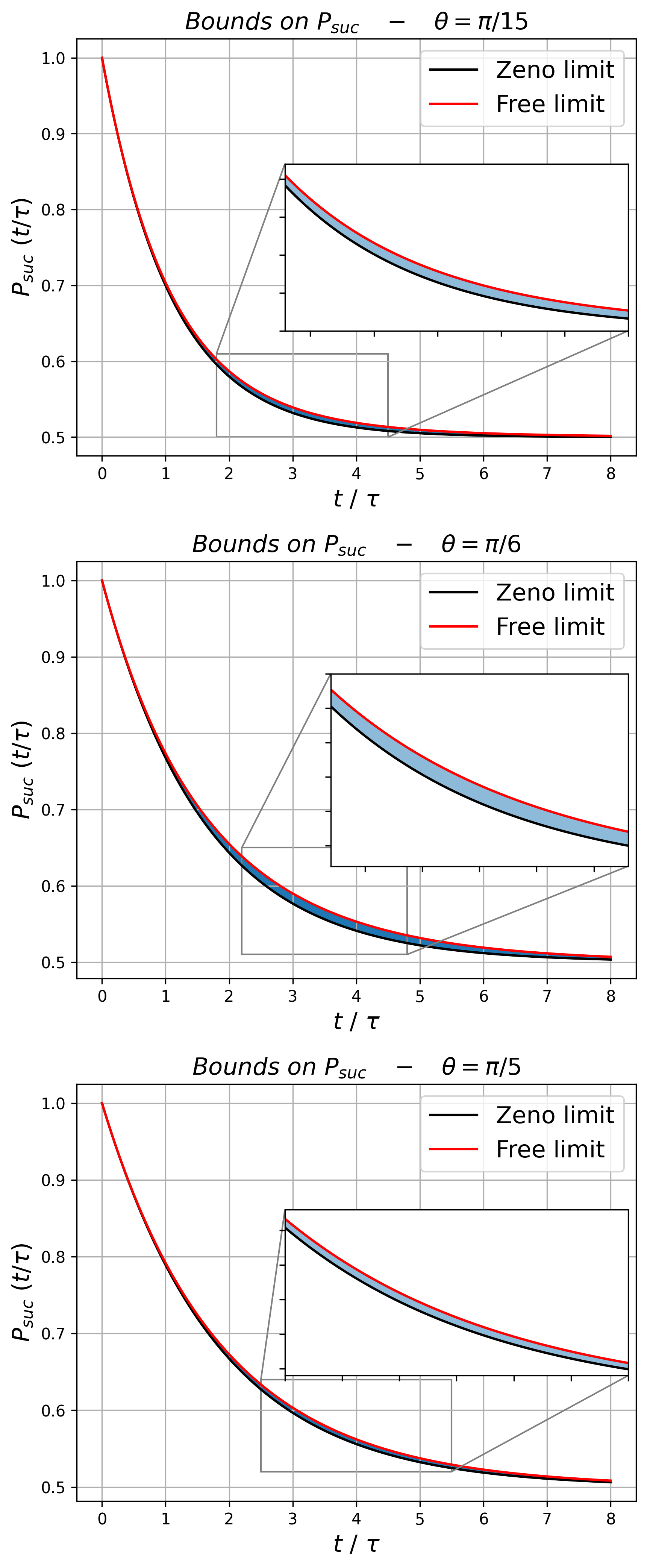}
    \caption{Success probability as a function of time in the Zeno and Free limits. Since the results are symmetrical with respect to $\pi/4$, the three angles were chosen to be distributed within the interval $0<\theta<\pi/4$, depicting the bounds in the limits $\theta\sim\pi/4$, $\theta\sim0$ and within an intermediate band. Any finite $\Delta$ will place the evolution in the shaded area between the two curves. These areas are better observed in the zoomed insets inside the graphs. The curves coincide when the angle $\theta$ is $0$, $\pi/4$ or $\pi/2$.}
    \label{fig:psuc}
\end{figure}

Interestingly, as we assume both initial states to be equally likely, the expressions for the success probability do not depend on the pointer basis ground state population $p$, and therefore do not depend on the temperature of the environment. In Fig. \ref{fig:psuc} we see that $P_{\mathrm{suc}}^{\mathrm{Zeno}}$ and $P_{\mathrm{suc}}^{\mathrm{Free}}$ provide, respectively, lower and upper bounds to the general $P_{\mathrm{suc}}$.

\section{\label{sec:conclusion}Conclusions}

We have shown that the quantum Zeno effect might be useful for preserving classical information stored in quantum systems even beyond the Zeno limit. After explicitly determining the effective dynamics for any time interval between operations, we have established bounds on the protocol's efficiency, as quantified by the probability of successfully storing classical information over a given period of time. This way, we have demonstrated the practical applicability and limitations of the quantum Zeno effect in our model. We expect these insights to contribute to the design and optimization of quantum information processing protocols, enabling better strategies for preserving information in the presence of noise. Further research in this direction may help advance the field of quantum information processing and pave the way for the realization of useful quantum computation in a near future.

\begin{acknowledgments}
GZ acknowledges financial support from FAPESP (Grant No. 2022/00993-9) and DOSP acknowledges the support from the funding agencies CNPq (Grant No. 307028/2019-4), FAPESP (Grant No. 2017/03727-0), and the Brazilian National Institute of Science and Technology of Quantum Information (INCT-IQ) Grant No. 465469/2014-0.
\end{acknowledgments}

\appendix

\section{\label{sec:app-ortho}Computational basis orthogonality constraint}

Using a parametrization reminiscent of the Bloch sphere for each vector of the computational basis we obtain
\begin{subequations}
    \begin{equation}
        \ket{0}=\cos\theta_0\ket{e_0}+e^{i\phi_0}\sin\theta_0\ket{e_1},
    \end{equation}
    \begin{equation}
        \ket{1}=\cos\theta_1\ket{e_0}+e^{i\phi_1}\sin\theta_1\ket{e_1},
    \end{equation}
\end{subequations}
for $\theta_0,\theta_1\in[0,\pi/2]$ and $\phi_0,\phi_1\in[0,2\pi)$. Despite being parameterized by four angles, they cannot vary freely as the orthogonality between $\ket{0}$ and $\ket{1}$ must be satisfied. Writing explicitly
\begin{equation}
    \begin{aligned}
    0=&\bra{0}\ket{1}\\
    =&\cos\theta_0\cos\theta_1+e^{i(\phi_1-\phi_0)}\sin\theta_0\sin\theta_1.
    \end{aligned}
\end{equation}
As the left hand side is real, so must be the right hand side, then
\begin{equation}
    e^{i(\phi_1-\phi_0)}=\pm 1,
\end{equation}
but since both $\sin$ and $\cos$ of $\theta_0$ and $\theta_1$ are positive in the considered intervals, we must have
\begin{equation}
    e^{i(\phi_1-\phi_0)}=-1,
\end{equation}
as it yields
\begin{equation}
    \sin\theta_0\sin\theta_1=\cos\theta_0\cos\theta_1.
\end{equation}
Therefore, we have
\begin{equation}
    e^{i\phi_0}=-e^{i\phi_1}:=e^{i\phi}.
\end{equation}
 Also, 
\begin{equation}
    \tan\theta_0\tan\theta_1=1,
\end{equation}
which in the considered intervals imply
\begin{equation}
    \sin\theta_1=\cos\theta_0:=\cos\theta,
\end{equation}
yielding the expressions as presented in the main text.

\section{\label{sec:app-gamma}Exponential decay}

Given a general initial state for the qubit
\begin{equation}
    \rho_{\mathrm{in}}=
    \begin{bmatrix}
    a & b \\
    b^* & 1-a
    \end{bmatrix}
\end{equation}
the effects of the Kraus operators are calculated from Eq. \eqref{eq:kraus}:
\begin{subequations}
    \begin{equation}
        K_0\rho_{\mathrm{in}}K_0^\dagger=p
    \begin{bmatrix}
    a & b\sqrt{1-\gamma} \\
    b^*\sqrt{1-\gamma} & 1-a(1-\gamma)
    \end{bmatrix},
    \end{equation}
    \begin{equation}
        K_1\rho_{\mathrm{in}}K_1^\dagger=p\gamma
    \begin{bmatrix}
    0 & 0 \\
    0 & 1-a
    \end{bmatrix},
    \end{equation}
    \begin{equation}
        K_2\rho_{\mathrm{in}}K_2^\dagger=(1-p)
    \begin{bmatrix}
    a(1-\gamma) & b\sqrt{1-\gamma} \\
    b^*\sqrt{1-\gamma} & 1-a
    \end{bmatrix},
    \end{equation}
    \begin{equation}
        K_3\rho_{\mathrm{in}}K_3^\dagger=(1-p)\gamma
    \begin{bmatrix}
    a & 0 \\
    0 & 0
    \end{bmatrix},
    \end{equation}
\end{subequations}
and from Eq. \eqref{eq:operator-sum}
\begin{equation}
    \begin{aligned}
        \rho_t=&~\mathcal{E}_t(\rho_{\mathrm{in}})\\
    =&    
    \begin{bmatrix}
    p-(1-\gamma_t)(p-a) & b\sqrt{1-\gamma_t} \\
    b^*\sqrt{1-\gamma_t} & (1-p)+(1-\gamma_t)(p-a)
    \end{bmatrix}.
    \end{aligned}
\end{equation}

Consider that we take the state $\rho_t$ and apply $\mathcal{E}_{t^\prime}$. Using the last result, we can set $a\to p-(1-\gamma_t)(p-a)$ and obtain for the $\ket{e_0}\bra{e_0}$ term
\begin{equation}
    \begin{aligned}
        \bra{e_0}\qty[\mathcal{E}_{t^\prime}(\rho_t)]\ket{e_0}=&~p-(1-\gamma_{t^\prime})(p-[p-(1-\gamma_t)(p-a)])\\
        =&~p-(1-\gamma_t)(1-\gamma_{t^\prime})(p-a)\\
        =&~p-f_tf_{t^\prime}(p-a),
    \end{aligned}
\end{equation}
where we defined
\begin{equation}
    f_t:=1-\gamma_t,
\end{equation}
which leads to
\begin{equation}
    \begin{aligned}
    \mathcal{E}_{t^\prime}(\rho_t)=&    
    \begin{bmatrix}
    p-f_tf_{t^\prime}(p-a) & b\sqrt{f_tf_{t^\prime}} \\
    b^*\sqrt{f_tf_{t^\prime}} & (1-p)+f_tf_{t^\prime}(p-a)
    \end{bmatrix}.
    \end{aligned}
\end{equation}

On the other hand, if we apply $\mathcal{E}_{t+t^\prime}$ on $\rho_{\mathrm{in}}$ we obtain
\begin{equation}
    \begin{aligned}
    \mathcal{E}_{t+t^\prime}(\rho_{\mathrm{in}})=&    
    \begin{bmatrix}
    p-f_{t+t^\prime}(p-a) & b\sqrt{f_{t+t^\prime}} \\
    b^*\sqrt{f_{t+t^\prime}} & (1-p)+f_{t+t^\prime}(p-a)
    \end{bmatrix}.
    \end{aligned}
\end{equation}

Notice that requiring Eq. \eqref{eq:homogeneous} to hold implies
\begin{equation}
    \mathcal{E}_{t^\prime}\qty[\mathcal{E}_{t}(\rho_{\mathrm{in}})]=\mathcal{E}_{t+t^\prime}(\rho_{\mathrm{in}})
\end{equation}
for any $\rho_{\mathrm{in}}$, meaning that for any $a,b\in\mathbb{C}$ we should have
\begin{equation}
    \begin{aligned}
    &\begin{bmatrix}
    p-f_{t+t^\prime}(p-a) & b\sqrt{f_{t+t^\prime}} \\
    b^*\sqrt{f_{t+t^\prime}} & (1-p)+f_{t+t^\prime}(p-a)
    \end{bmatrix}
    =\\
    &\begin{bmatrix}
    p-f_tf_{t^\prime}(p-a) & b\sqrt{f_tf_{t^\prime}} \\
    b^*\sqrt{f_tf_{t^\prime}} & (1-p)+f_tf_{t^\prime}(p-a)
    \end{bmatrix},
    \end{aligned}
\end{equation}
which is equivalent to the condition
\begin{equation}
    f_tf_{t^\prime}=f_{t+t^\prime}.
\end{equation}

Now, we show that the only function that satisfies this condition is the exponential. Consider the derivative
\begin{equation}
    \begin{aligned}
        f^\prime_t=&\lim_{h\to0}\frac{f_{t+h}-f_t}{h}\\
        =&\lim_{h\to0}\frac{f_tf_h-f_t}{h}\\
        =&~f_t\lim_{h\to0}\frac{f_h-1}{h}.
    \end{aligned}
\end{equation}
The limit only exists if we further assume $\lim_{t\to0}f_t=1$, or simply $f_0=1$ for a continuous function, in which case the L'Hôpital rule applies
\begin{equation}
    \begin{aligned}
        f^\prime_t=&~f_t\lim_{h\to0}\frac{\dv{}{h}[f_h-1]}{\dv{}{h}h}\\
        =&~f_t\lim_{h\to0}f^\prime_h\\
        =&~f_tf^\prime_0.
    \end{aligned}
\end{equation}
This simple differential equation has solutions of the form
\begin{equation}
    f_t=e^{\alpha t},~\alpha\in\mathbb{R},
\end{equation}
while returning to $\gamma_t$ yields
\begin{equation}
    \gamma_t=1-e^{\alpha t}.
\end{equation}
Given that by hypothesis we must have $\dot{\gamma}>0$, then $\alpha$ is necessarily negative, which allows us to define a positive characteristic time for the dynamics
\begin{equation}
    \tau:=-\frac{1}{\alpha},
\end{equation}
finally obtaining
\begin{equation}
    \gamma_t=1-e^{-\frac{t}{\tau}}.
\end{equation}

\section{\label{sec:app-ground}Ground state population}

We simplify Eq. \eqref{eq:recurrence} by defining
\begin{equation}
    \alpha:=P_{0|1},
\end{equation}
\begin{equation}
    \beta:=1-\qty(P_{0|1}+P_{1|0}),
\end{equation}
and rewrite it as a discrete evolution
\begin{equation}
    a_{n+1}=\alpha+\beta a_n.
\end{equation}

Notice that expanding for a few terms yields
\begin{subequations}
    \begin{equation}
        a_1=\alpha+\beta a_0,
    \end{equation}
    \begin{equation}
        a_2=\alpha+\alpha\beta+\beta^2 a_0,
    \end{equation}
    \begin{equation}
        a_3=\alpha+\alpha\beta+\alpha\beta^2+\beta^3 a_0,
    \end{equation}
\end{subequations}
so for a general term we have
\begin{equation}
    a_n=\alpha \sum_{i=0}^{n-1}\beta^i + \beta^n a_0,
\end{equation}
and using the known result for the sum of a geometrical progression
\begin{equation}
    \begin{aligned}
        a_n=&~\alpha \frac{\beta^n-1}{\beta-1} + \beta^n a_0\\
        =&~\frac{\alpha}{1-\beta}-\qty(\frac{\alpha}{1-\beta}-a_0)\beta^n.
    \end{aligned}
\end{equation}
Then we notice that
\begin{equation}
    \frac{\alpha}{1-\beta}=\frac{P_{0|1}}{P_{0|1}+P_{1|0}}=a_{\infty}
\end{equation}
and return to the original variables
\begin{equation}
    a(t)=a_{\infty}-\qty(a_{\infty}-a_0)\qty[1-\qty(P_{0|1}+P_{1|0})]^{t/\Delta}.
\end{equation}

\section{\label{sec:app-trans}Transition probabilities}

For simplicity, we calculate $P_{0|0}$ from the Kraus representation of the amplitude damping channel and use $P_{1|0}=1-P_{0|0}$,
\begin{equation}
\label{eq:channel}
    \begin{aligned}
    P_{0|0}=&\bra{0}\rho_{\Delta}\ket{0}\\
    =&\bra{0}\qty[\sum_{i=0}^3K_i\rho_{\mathrm{in}}K_i^\dagger]\ket{0}\\
    =&\bra{0}\qty[\sum_{i=0}^3K_i\qty(\ket{0}\bra{0})K_i^\dagger]\ket{0}\\
    =&\sum_{i=0}^3\bra{0}K_i\ket{0}\bra{0}K_i^\dagger\ket{0},
    \end{aligned}
\end{equation}
requiring the calculation of $\bra{0}K_i\ket{0}$ for each $K_i$. Eq. \eqref{eq:kraus} yields
\begin{subequations}
    \begin{equation}
        \bra{0}K_0\ket{0}=\sqrt{p}\qty(\cos^2\theta+\sqrt{1-\gamma}\sin^2\theta),
    \end{equation}
    \begin{equation}
        \bra{0}K_1\ket{0}=\sqrt{p}\sqrt{\gamma}\cos\theta\sin\theta,
    \end{equation}
    \begin{equation}
        \bra{0}K_2\ket{0}=\sqrt{1-p}\qty(\sqrt{1-\gamma}\cos^2\theta+\sin^2\theta),
    \end{equation}
    \begin{equation}
        \bra{0}K_3\ket{0}=\sqrt{1-p}\sqrt{\gamma}\cos\theta\sin\theta,
    \end{equation}
\end{subequations}
and notice that in all the four cases $\bra{0}K_i^\dagger\ket{0}=\bra{0}K_i\ket{0}$, implying that $\bra{0}K_i\ket{0}\bra{0}K_i^\dagger\ket{0}=\qty(\bra{0}K_i\ket{0})^2$. These results may be inserted into Eq. \eqref{eq:channel}
\begin{equation}
    \begin{aligned}
    P_{0|0}=& \sum_{i=0}^3\qty(\bra{0}K_i\ket{0})^2\\
    =&~p\qty(\cos^2\theta+\sqrt{1-\gamma}\sin^2\theta)^2\\
    +&~p\gamma\cos^2\theta\sin^2\theta\\
    +&~(1-p)\qty(\sqrt{1-\gamma}\cos^2\theta+\sin^2\theta)^2\\
    +&~(1-p)\gamma\cos^2\theta\sin^2\theta\\
    =&~\qty(1-\gamma+p\gamma)\cos^4\theta+\qty(1-p\gamma)\sin^4\theta\\
    +&~\qty(\gamma+2\sqrt{1-\gamma})\cos^2\theta\sin^2\theta,
    \end{aligned}
\end{equation}
and applying trigonometric identities results in
\begin{equation}
    \begin{aligned}
    \label{eq:p00}
    P_{0|0}=&~ \frac{1}{4}\Big[3-\gamma+\sqrt{1-\gamma}+2\gamma(2p-1)\cos(2\theta)\\
    +&~\qty(1-\gamma-\sqrt{1-\gamma})\cos(4\theta)\Big],
    \end{aligned}
\end{equation}
while similar calculations for $P_{1|1}$ yield
\begin{equation}
    \begin{aligned}
    \label{eq:p11}
    P_{1|1}=&~ \frac{1}{4}\Big[3-\gamma+\sqrt{1-\gamma}-2\gamma(2p-1)\cos(2\theta)\\
    +&~\qty(1-\gamma-\sqrt{1-\gamma})\cos(4\theta)\Big].
    \end{aligned}
\end{equation}

Finally, we use the explicit expression for $\gamma(t)$ and return to the transition probabilities
\begin{subequations}
\label{eq:trans}
    \begin{equation}
        \begin{aligned}
            P_{0|1}=&~\frac{1}{4}\Bigg[\Big(1-\cos(4\theta)\Big)\qty(1-e^{-\Delta/2\tau})\\
            +&~\Big(1+2(2p-1)\cos(2\theta)+\cos(4\theta)\Big)\qty(1-e^{-\Delta/\tau})\Bigg]\\
        \end{aligned}
    \end{equation}
    \begin{equation}
        \begin{aligned}
            P_{1|0}=&~\frac{1}{4}\Bigg[\Big(1-\cos(4\theta)\Big)\qty(1-e^{-\Delta/2\tau})\\
            +&~\Big(1-2(2p-1)\cos(2\theta)+\cos(4\theta)\Big)\qty(1-e^{-\Delta/\tau})\Bigg].
        \end{aligned}
    \end{equation}
\end{subequations}

Now we can explicitly write
\begin{equation}
    \begin{aligned}
            P_{0|1}+P_{1|0}=&~1-\frac{1}{2}\Big[\qty(1-\cos(4\theta))e^{-\Delta/2\tau}\\
            +&~\qty(1+\cos(4\theta))e^{-\Delta/\tau}\Big]\\
            =&~1-\qty[\sin^2(2\theta)e^{-\Delta/2\tau}+\cos^2(2\theta)e^{-\Delta/\tau}]
        \end{aligned}
\end{equation}
and
\begin{equation}
    a_{\infty}(\Delta,\theta)=\frac{1}{2}\qty[1+\frac{(2p-1)\cos(2\theta)}{\cos^2(2\theta)+\sin^2(2\theta)\frac{1-e^{-\Delta/2\tau}}{1-e^{-\Delta/\tau}}}],
\end{equation}
as shown in the main text.

\end{document}